\documentstyle[11pt,epsfig]{article}
\title{\bf The modulated soliton fields in the Goldstone boson model}
\author {Z.Shahbazi$^1$, S. Miraboutalebi$^1$ and F. Ahmadi$^2$\footnote{fahmadi@sru.ac.ir}\\
$^1${\small Department of Physics, Islamic Azad University, North Tehran Branch, Tehran, 1651153311, Iran.}\\
$^2${\small Department of Physics, Shahid Rajaee Teacher Training University, Tehran 1678815811, Iran.}\\}

\begin{document}
\maketitle
\begin{abstract}
It is well known that massless Goldstone bosons have not yet been observed. In the Goldstone boson model, after the spontaneous symmetry breaking under $U(1)$,
two coupled nonlinear equations are obtained for which we present the exact solitonic solutions. These solutions completely localize
the energy density of the model and show the existence of two boson fields, one massive and the other massless. Also, it is seen that the solitonic waves are modulated and the massless wave rides on the massive wave. Finally, we calculate the charge density of the model, which again confirms the neutrality of these boson fields. Since massive bosons can be observed in the laboratory, these solitonic waves may be useful in tracking and detecting Goldstone bosons.
\end{abstract}


%%%%%%%%%%%%%%%%%%%%%%%%%%%%%%%%%%%%%%%%%%%%%%%%%%%%%%%%%%%%%%%%%%%%%%%%%%%%%%%

\section{Introduction}
Symmetry plays an important role in our understanding of nature. Since the birth of science endeavor, symmetry structures in nature have attracted the minds of natural philosophers and physicists, looking for the origin of the laws of nature \cite{Bra}. In some theories, spontaneous symmetry breaking (SSB) happens when the ground state does not respect the symmetry of the theory \cite{Kib}. One such theory is that of the Goldstone boson model with a degenerate ground state. The symmetry of the model spontaneously breaks under continuous transformation $\phi(x,t)\rightarrow \phi(x,t)e^{i\alpha}$ and  leads to the appearance of two spin-zero particles, a massless and a massive boson. The massless boson is known as the Nambu-Goldstone boson \cite{Mnd}. Since massless bosons have not any reasonable interaction strength, they cannot be observed easily, but the massless particles (pseudo-Nambu-Goldstone bosons) have many laboratory, astrophysical and cosmological implications \cite{Ball, Luz}. %play an important role in explanation of cosmological phenomena.
For example, they are considered as the generator of cosmological dark energy potential which causes present cosmic acceleration, so expansion of the universe can result from such particles \cite{Ad,Ki}.

There are numerous  reviews of the physics of Nambu-Goldstone bosons and  their effective field theory \cite{Geo}-\cite{Bur}. The Goldstone boson model is the simplest model to undergo SSB which involves a complex scalar field with a self-interaction term described by the potential function ${\cal V}(\phi)=\mu^{2}|\phi(x)|^{2}+\lambda|\phi(x)|^{4}$. Regardless of the $\lambda|\phi(x)|^{4}$ term that $(\lambda>0)$, the govering equation of the model is the same as complex Klein-Gordon equation. In the Goldstone boson model after SSB, the complex field can be defined by two real fields which are derivatives around the ground state, so the Lagrangian of the model is defined according the fields that lead to two coupled nonlinear equations. The equations have already been treated by perturbation methods but suffer from problems such as inequality of the lagrangian with the initial lagrangian. The equivalence of Lagrangians holds only for exact solutions of the theory \cite{Alon, Kan}.
%and calculations about an unstable solution, that can not be done in the perturbation theory \cite{Alon, Kan}. On the other hand, the quantized theory shows that of the unperturbed system corresponding to particles of imaginary mass for which no order of perturbation theory can offer an explanation right \cite{Pere, Arkani, Ski}.
In this work, the exact solutions of the coupled nonlinear equations are introduced by using  $\tanh/\mbox{sech}$  method for which soliton solutions appear. %which do not suffer from the above  mentioned problems.
Soliton solutions of the nonlinear partial differential equations (NPDEs) can illustrate many phenomena in physics, they are created as a result of a neutralization of the nonlinear and dispersive effects. The interesting feature of a soliton is that, after collision with another soliton it keeps its properties which is similar to that of a particles \cite{A}.

 The $\tanh$ method is a powerful technique to obtain traveling wave solutions \cite{Mal1}-\cite{Wazwa}. In other words, the method is appropriate
to solve equations where dispersion, convection and reaction diffusion phenomena play an important
role in such models \cite{Jaw}. For example the method is used to solve nonlinear Klein-Gordon equation \cite{Sir}-\cite{Waz2}, nonlinear fifth-order KDV equation \cite{Waz3} and nonlinear conduction and Burgers-Fisher equations \cite{Waz4}. The extended $\tanh$ method \cite{Waz1, Abd} and the modified extended $\tanh$-function method \cite{Sol, Wak} are the extended version of this approach.

In the Goldstone boson model to obtain exact solutions, at first we find the field equations in the form of two coupled (NPDEs), then by using the $\tanh$ method we convert the equations to nonlinear ordinary differential equations (NODEs). By solving the equations we get solitary solutions for which we choose the localized solutions. The soliton solutions are plotted and two modulated waves are observable in the figure for which the massless one rides on the massive one. Finally, the Hamiltonian density and charge density of the model are calculated. The figures of the energy density show that two localized fields exist in the model. It can also be seen that the charge density becomes zero, so the total charge and the electric dipole moment of the model become zero, which corresponds to the neutrality of the produced bosons. Since massive bosons can be observed in the laboratory, these solitonic waves may be useful in designing experiments to detect massless bosons. So, these studies could impact our understanding of dark matter, the problem of $CP$ violation and the hierarchy problem \cite{Kimball}.
%It can also be seen that the continuity equation for the charge-current density of these fields is also satisfied. Interestingly, the total electric charge in the model is still zero, but the electric dipole moment becomes non-zero. In fact, the spontaneous breaking of the symmetry causes polarization to occur in the model. It seems that this electric polarization and non-zero electric dipole moment are very useful in designing experiments to detect massless bosons and to investigate the problem of $CP$ violation \cite{Luce}. So, these studies could impact our understanding of the matter-antimatter asymmetry of the Universe \cite{Ball}.

This work organized as follows. In section two, the model is briefly presented and the field equations are derived. Also, the Hamiltonian density and charge-current density of the model are written. In section three, using the $\tanh/\mbox{sech}$ method, the soliton solutions are obtained and the figure of these modulated waves is plotted. In the first subsection of this section, the Hamiltonian density is determined which its figures show the localized energy density and two peaks close to each other. In the second subsection, the charge-current density of the model is calculated. Finally the conclusion are drawn in the last section. The relations and the model are written in natural units, namely, in $\hbar=c=m_{e}=\epsilon_{0}=1$ and the metric signature is $(-, +, +, +)$.

%%%%%%%%%%%%%%%%%%%%%%%%%%%%%%%%%%%%%%%%%%%%%%%%%%%%%%%%%%%%%%%%%%%%%%%%%%%%%%%%%%%%%%%%%%%%%%%%%%%%%%%%%%%%%%%%%%%%%%%%%%%%%%%%%%%%%%%%%%%%%%%%%%%%%%%%%%%%%%%%
\section{The Model}
The model is described by the Lagrangian density
\begin{equation}\label{Ia}
{\cal L}=\partial^{\alpha}\phi^{\ast}(x^{\beta})\partial_{\alpha}\phi(x^{\beta})-\mu^{2}|\phi(x^{\beta})|^{2}-\lambda|\phi(x^{\beta})|^{4}\,,\,\,\,\,\ (\lambda>0)\,,
\end{equation}
where $\phi(x)$ is a classical complex scalar field, $\mu^{2}$ and $\lambda$ are arbitrary real parameters. It is obvious the Lagrangian density without interaction term $\lambda|\phi(x^{\beta})|^{4}$ is the same as the free complex Klein-Gordon field. The Lagrangian density is invariant under the following global transformations \cite{Mnd}
\begin{equation}\label{Ic}
\phi(x^{\beta})\longrightarrow \acute{\phi}(x^{\beta})=\phi(x^{\beta}) e^{i\gamma}\,, \hspace{1cm}  \phi^{\ast}(x^{\beta})\longrightarrow \phi^{\acute{\ast}}(x^{\beta})=\phi^{\ast}(x) e^{-i\gamma}\,.
\end{equation}
In order to find the ground state, minimum of the potential density must be obtained, so two different situations occur, depending on the sign of $\mu^{2}$. The potential density at $\phi=0$ has a absolute minimum for $\mu^{2}>0$, it is clear that the (SSB) can not occur in this situation. For $\mu^{2}<0$, the potential density has a maximum at $\phi=0$ and a set of absolute minimum at $\phi(x)=\phi_{0}=(\frac{-\mu^{2}}{2\lambda})^{1/2}e^{i\theta}\,,\,\ 0\leq \theta<2\pi$. Since the Eq.(\ref{Ia}) is invariance under the global phase transformation, the value of $\theta$ chosen is not important and it can be taken zero, therefore
\begin{equation}\label{Ig}
\phi_{0}=\left(\frac{-\mu^{2}}{2\lambda}\right)^{1/2}= \frac{1}{\sqrt{2}}\,\ \nu \,,\,\,\,\,\ (\nu>0)\,,
\end{equation}
where $\nu=(\frac{-\mu^{2}}{\lambda})^{\frac{1}{2}}$ and $\phi_{0}$ is real. It can be seen  the (SSB) occurs by choosing one particular direction. The field $\phi(x^{\beta})$ around the minimum of the potential can be written as follows
\begin{equation}\label{Ih}
\phi(x^{\beta})=\frac{1}{\sqrt{2}}\left[\nu+\psi_{1}(x^{\beta})+i\psi_{2}(x^{\beta})\right]\,,
\end{equation}
where $\psi_{1}(x^{\beta})$ and $\psi_{2}(x^{\beta})$ are real and measure the deviations of the field $\phi(x^{\beta})$ from the equilibrium ground state $\phi_{0}$. In turns of $\psi_{1}(x^{\beta})$ and $\psi_{2}(x^{\beta})$ the Lagrangian density become
\begin{eqnarray}
{\cal L}&=&\frac{1}{2}\partial^\alpha\psi_{1}(x^{\beta})\partial_{\alpha}\psi_{1}(x^{\beta})-\lambda\nu^{2}\psi_{1}^{2}(x^{\beta})
+\frac{1}{2}\partial^\alpha\psi_{2}(x^{\beta})\partial_{\alpha}\psi_{2}(x^{\beta})-\lambda\nu\psi_{1}(x^{\beta}) \left[\psi_{1}^{2}(x^{\beta})+\psi_{2}^{2}(x^{\beta})\right]
\nonumber\\&-&\frac{1}{4}\lambda\left[\psi_{1}^{2}(x^{\beta})+\psi_{2}^{2}(x^{\beta})\right]^{2}\,.\label{Ii}
\end{eqnarray}
According to the above Lagrangian, it is clear that the fields $\psi_{1}(x^{\beta})$ and $\psi_{2}(x^{\beta})$ are real Klein-Gordon fields and in the quantization process, both fields lead to neutral spin $0$ particles that the $\psi_{1}$ boson with the mass $(\sqrt{2\lambda\nu^{2}})$ and the $\psi_{2}$ boson which has zero mass \cite{Mnd}. On the other hands, Eq.(\ref{Ia}) and Eq.(\ref{Ii}) are the same Lagrangian density expressed in terms of different variables. So, they are entirely equivalent and must lead to the same physical results. This equivalence only holds for exact solutions of the theory, but in perturbation theory, approximate solutions lead to a very different result \cite{Mnd}. In this work, we find the exact solutions for the above Lagrangian density and can get interesting information about the fields. Applying the Euler-Lagrange equation to the Lagrangian density (\ref{Ii}), we get the following relations
\begin{equation}\label{Ij}
\frac{1}{2}\Box\psi_{1}(x^{\beta})+2\lambda\nu^{2}\psi_{1}(x^{\beta})+3\lambda\nu\psi_{1}^{2}(x^{\beta})
+\lambda\nu\psi_{2}^{2}(x^{\beta})+\lambda\psi_{1}(x^{\beta})\left[\psi_{1}^{2}(x^{\beta})+\psi_{2}^{2}(x^{\beta})\right]=0\,,
\end{equation}
and
\begin{equation}\label{Ik}
\frac{1}{2}\Box\psi_{2}(x^{\beta})+2\lambda\nu\psi_{1}(x^{\beta})\psi_{2}(x^{\beta})+\lambda\psi_{2}(x^{\beta})\left[\psi_{1}^{2}(x^{\beta})+\psi_{2}^{2}(x^{\beta})\right]=0\,,
\end{equation}
since the above equations are coupled nonlinear equations, to get their exact solutions we use the $\tanh$ approach and obtain the soliton solutions in the next section. For energy density of the model, we calculate the following Hamiltonian density
\begin{eqnarray}
{\cal H}&=&\frac{1}{2}\left[\dot{\psi_{1}}^{2}(x^{\beta})+\dot{\psi_{2}}^{2}(x^{\beta})+(\nabla\psi_{1}(x^{\beta}))^{2}+(\nabla\psi_{2}(x^{\beta}))^{2}\right]
+\lambda\nu^{2}\psi_{1}^{2}(x^{\beta})+\lambda\nu\psi_{1}(x^{\beta})
\nonumber\\&\times&\left[\psi_{1}^{2}(x^{\beta})+\psi_{2}^{2}(x^{\beta})\right]
+\frac{\lambda}{4}\left[\psi_{1}^{2}(x^{\beta})+\psi_{2}^{2}(x^{\beta})\right]^{2}\,,\label{Il}
\end{eqnarray}
and for charge-current density, using the definition (\ref{Ih}), we have
\begin{equation}
J_{\alpha}(x^{\beta})= q\left[-\nu \partial_{\alpha}\psi_{2}(x^{\beta})-\partial_{\alpha}\psi_{2}(x^{\beta}) \psi_{1}(x^{\beta})+ \partial_{\alpha}\psi_{1}(x^{\beta})\psi_{2}(x^{\beta})\right]\,,\label{Ila}
\end{equation}
which obviously satisfies the continuity equation, $\partial^{\alpha}J_{\alpha}(x^{\beta})=0$.
%%%%%%%%%%%%%%%%%%%%%%%%%%%%%%%%%%%%%%%%%%%%%%%%%%%%%%%%%%%%%%%%%%%%%%%%%%%%%%%%%%%%%%%%%%%%%%%%%%%%%%
\section{Soliton solutions}
In this section, we are going to solve the coupled nonlinear Eqs.(\ref{Ij}) and (\ref{Ik}) in one dimension using the $\tanh$ approach. For this purpose, we first consider the following variable
\begin{equation}\label{La}
\xi=g(x-wt) \,,
\end{equation}
where $w$ is velocity of the traveling waves and $g$ is wave number, that its inverse is proportional to width of the traveling waves. Then we replace $\psi_{1}$ and $\psi_{2}$ with
\begin{equation}\label{Lp}
\psi_{1}(x,t)=V(\xi), \hspace{1cm}  \psi_{2}(x,t)=U(\xi)\,,
\end{equation}
where $V(\xi)$ and $U(\xi)$ will be the solitary solutions for the massive and massless fields respectively.
Any static localized solution of the equations is a solitary wave which it can be transformed to a moving coordinate frame by a simple boost, so an advanced mode $(x+wt)$ or a retarded mode $(x-wt)$ is obtained and because of the nonlinearity, a simple linear combination of both of these modes does not provide a proper solution \cite{Jah, Raj}.
%After performing the first step, namely changing the coordinates Eq.(\ref{La}), we introduce $z=sech(\xi)$ as a new dependent variable. Because in this way, getting normalized and localized solutions is more possible comparing using $z=tanh(\xi)$ variable.
%\subsection{Soliton solutions}
By considering the above definitions, the Eqs.(\ref{Ij}) and (\ref{Ik}) become
\begin{equation}\label{Lb}
\frac{\eta}{2}\frac{d^{2}V}{d\xi^{2}}+2\lambda\nu^{2} V+3\lambda\nu V^{2}+\lambda\nu U^{2}+\lambda V[V^{2}+U^{2}]=0\,,
\end{equation}
and
\begin{equation}\label{Lc}
\frac{\eta}{2}\frac{d^{2}U}{d\xi^{2}}+2\lambda\nu V U +\lambda U[V^{2}+U^{2}]=0\,,
\end{equation}
where $\eta$ is a constant defined as
\begin{equation}\label{Ld}
\eta=g^{2}\left(1-w^{2}\right)\,.
\end{equation}
 To display the solitary solutions as polynomials of $\mbox{sech}$ function, the following series is supposed
\begin{equation}\label{Le}
V(\xi)=\emph{\textbf{V}}(Z)=\sum_{m=0}^{M}b_{m}Z^{m}\,,
\end{equation}
and
\begin{equation}\label{Lf}
U(\xi)=\emph{\textbf{U}}(Z)=\sum_{n=0}^{N}a_{n}Z^{n}\,,
\end{equation}
where $Z=\mbox{sech}(\xi)$.
By putting the above series in  Eqs.(\ref{Lb}) and (\ref{Lc}), we have
% Considering the following series the above equations and $z=sech(\xi)$, all the derivatives of the Eqs.(\ref{Lb}) and (\ref{Lc}) get changed and the equations become
\begin{eqnarray}
&&-\frac{\eta}{2}\left(\frac{d^{2}\emph{\textbf{V}}(Z)}{dZ^{2}}\right) Z^{4}-\eta\left(\frac{d \emph{\textbf{V}}(Z)}{dZ}\right) Z^{3}+\eta \left(\frac{d^{2}\emph{\textbf{V}}(Z)}{dZ^{2}}\right) Z^{2}
+\frac{\eta}{2}\left(\frac{d \emph{\textbf{V}}(Z)}{dZ}\right) Z+2\lambda\nu^{2}\emph{\textbf{V}}(Z)\nonumber\\&+&3\lambda\nu \emph{\textbf{V}}^{2}(Z)
+\lambda\nu \emph{\textbf{U}}^{2}(Z)+\lambda \emph{\textbf{V}}(Z)\emph{\textbf{U}}^{2}(Z)+\lambda \emph{\textbf{V}}^{3}(Z)=0\,,\label{Lg}
\end{eqnarray}
and
\begin{eqnarray}
&&-\frac{\eta}{2}\left(\frac{d^{2}\emph{\textbf{U}}(Z)}{dZ^{2}}\right) Z^{4}-\eta\left(\frac{d \emph{\textbf{U}}(Z)}{dZ}\right) Z^{3}+\eta \left(\frac{d^{2}\emph{\textbf{U}}(Z)}{dZ^{2}}\right) Z^{2}
+\frac{\eta}{2}\left(\frac{d \emph{\textbf{U}}(Z)}{dZ}\right) Z
+2\lambda \nu \emph{\textbf{V}}(Z)\emph{\textbf{U}}(Z)\nonumber\\&+&\lambda \emph{\textbf{U}}(Z)\emph{\textbf{V}}^{2}(Z)+\lambda \emph{\textbf{U}}^{3}(Z)=0\,.\label{Lh}
\end{eqnarray}
Now, considering Eqs.(\ref{Lg}) and (\ref{Lh}), parameters $N$ and $M$ in Eq.(\ref{Le}) and (\ref{Lf}) can be found by balancing the highest-order linear term with the highest-order nonlinear term \cite{Mal1}. It turns out that $N=M=1$, therefore we find
\begin{equation}\label{Ll}
\emph{\textbf{V}}(Z)=b_{0}+b_{1}Z\,,
\end{equation}
and
\begin{equation}\label{Lm}
\emph{\textbf{U}}(Z)=a_{0}+a_{1}Z\,,
\end{equation}
where $b_{0}$, $b_{1}$, $a_{0}$, $a_{1}$ are four unknown parameters. Then, by inserting Eqs.(\ref{Ll}) and (\ref{Lm}) in the Eqs.(\ref{Lg}) and (\ref{Lh}),
some equations of $Z = \mbox{sech}(\xi)$ are obtained. In order to hold the equations for any $Z$, the coefficients for
each power of $Z$ should be zero. In this way, eight coupled equations of the unknown parameters
of the theory are obtained. Solving these equations, an acceptable set of parameters can be found as follows
%We choose solutions for the fields which satisfy the normalization condition, therefore we consider the following relation
%\begin{equation}\label{Lx}
%\int_{-\infty}^{+\infty}d\xi|{\textbf{V}(\xi)}|^2=1   \hspace{1cm} and  \hspace{1cm}  \int_{-\infty}^{+\infty}d\xi|{\textbf{U}(\xi)}|^2=1\,.
%\end{equation}
%and for the massless field we have
%\begin{equation}\label{Ly}
%N=\int_{-\infty}^{+\infty}d\xi|{\textbf{U}(\xi)}|^2=1\,.
%\end{equation}
%By solving  the above eight coupled equations, we get an acceptable set of solutions as follows
\begin{equation}\label{Ln}
a_{0}=0\,,\,\,\  a_{1}=\pm\sqrt{2 b_{0}^{2}-b_{1}^{2}}\ \ \ \ (2 b_{0}^{2}-b_{1}^{2}>0) \,,\,\,\ b_{0}=b_{0} \,,\,\,\ b_{1}=b_{1} \,,\,\,\ \lambda=\frac{\eta}{2b_{0}^{2}} \,,\,\,\ \nu=-b_{0}\,,
\end{equation}
where $b_{1}$ is free parameter but should satisfy the condition for $a_{1}$ in the above relation. Using above parameters, we get the soliton wave solutions of Eqs.(\ref{Ll}) and (\ref{Lm}) as follows
\begin{equation}\label{Lo}
U(\xi)=\sqrt{2 \nu^{2}-b_{1}^{2}}\,\, \mbox{sech}(\xi)\,,  \hspace{1cm}  V(\xi)= -\nu+b_{1}\mbox{sech}(\xi)\,.
\end{equation}
By substituting the above solutions in relation (\ref{Ih}), $\phi(x^{\beta})$ is obtained as a normalizable wave function for each value of parameter $b_{1}$.
Therefore, without losing the generality of the problem, the parameter $b_{1}$ can be chosen such that the massless field is normalized. These solutions are plotted in Fig.(1), which shows both modulated solitons and the massless field rids on the massive one. It is notice that these fields must lead to the localized energy density, to be suitable for the model. So, we explore this subject in the next section.
\begin{figure}[ht]
\begin{center}
 \includegraphics[width=6cm]{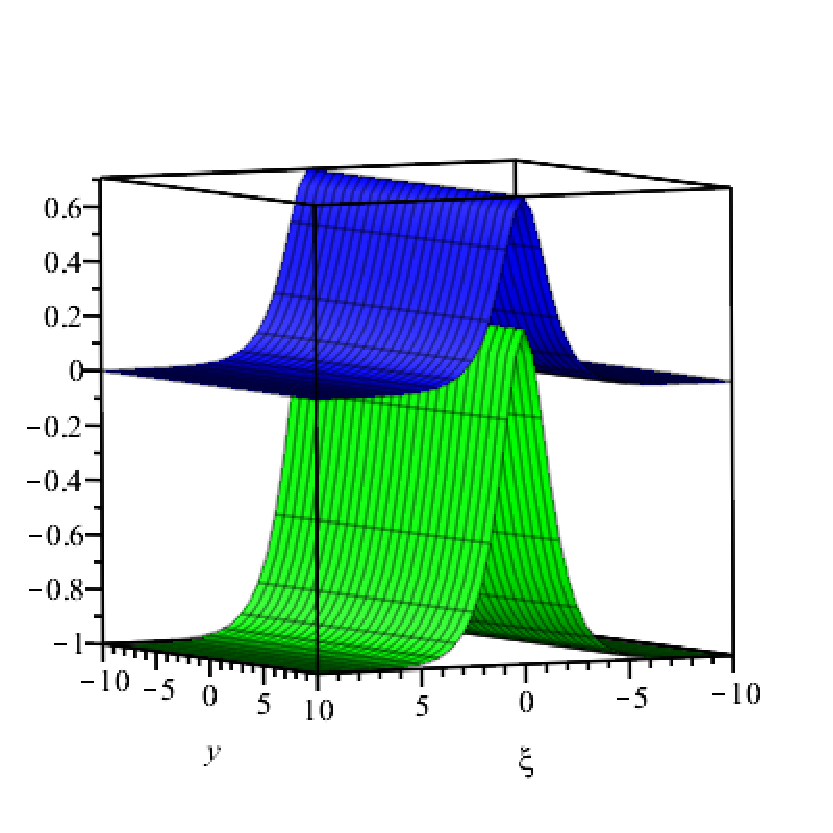}
\caption{\small{The fields are plotted for $\nu=1$ and $b_{1}=1.224$. The blue one is massless soliton and the green one is massive soliton.}}
\end{center}
\end{figure}

\subsection{ The energy density of the model}
For the energy density, using Eq.(\ref{Il}), we calculate the Hamiltonian density as follows
\begin{eqnarray}
{\cal H}(\xi, w, g)&=&\frac{1}{2}\left[\left(\frac{\partial \psi_{1}(\xi)}{\partial\xi}\frac{\partial \xi}{\partial t}\right)^{2}+\left(\frac{\partial \psi_{2}(\xi)}{\partial \xi}\frac{\partial \xi}{\partial t}\right)^{2}+\left(\frac{\partial \psi_{1}(\xi)}{\partial \xi}\frac{\partial \xi}{\partial x}\right)^{2}+\left(\frac{\partial \psi_{2}(\xi)}{\partial \xi}\frac{\partial \xi}{\partial x}\right)^{2}\right]
+\lambda \nu^{2}\psi_{1}^{2}(\xi)\nonumber\\&+&\lambda \nu \psi_{1}(\xi)\left[\psi_{1}^{2}(\xi)+ \psi_{2}^{2}(\xi)\right]+\frac{\lambda}{4}\left[\psi_{1}^{2}(\xi)+\psi_{2}^{2}(\xi\underline{})\right]^{2}\,,\label{Ls}
\end{eqnarray}
By putting the obtained fields (\ref{Lo}) in the above relationship, we have
\begin{eqnarray}
&&{\cal H}(\xi, w, g)=\frac{g^{2}(w^{2}+1)}{2}\left[\left( b_{1}\mbox{sech}(\xi)\tanh(\xi)\right)^{2}+\left(\sqrt{2\nu^{2}-b_{1}^{2}}\mbox{sech}(\xi)\tanh(\xi)\right)^{2}\right]
+\frac{g^{2}(1-w^{2})}{2}
\nonumber\\&\times&\left[-\nu+b_{1}\mbox{sech}(\xi)\right]^{2}
+\frac{g^{2}(1-w^{2})}{2\nu}\times  \left(-\nu+b_{1}\mbox{sech}(\xi)\right)\left[\left(-\nu+b_{1}\mbox{sech}(\xi)\right)^{2}+\left(\sqrt{2\nu^{2}-b_{1}^{2}}\mbox{sech}(\xi)\right)^{2}\right]
\nonumber\\&+&\frac{g^{2}(1-w^{2})}{8\nu^{2}}\left[\left(-\nu+b_{1}\mbox{sech}(\xi)\right)^{2}+\left(\sqrt{2\nu^{2}-b_{1}^{2}}\mbox{sech}(\xi)\right)^{2}\right]^{2}\,.\label{Lt}
\end{eqnarray}
The Hamiltonian density is plotted in the Fig.(2) which is localized, and since the soliton waves are modulated in the Fig.(1), two peaks close to each other are observable in the Fig.(2) that shows the existence of two fields. Also, to make this clear, Fig.(3) is plotted for a specific $g$.
 \begin{figure}[ht]
\begin{center}
 \includegraphics[width=6cm]{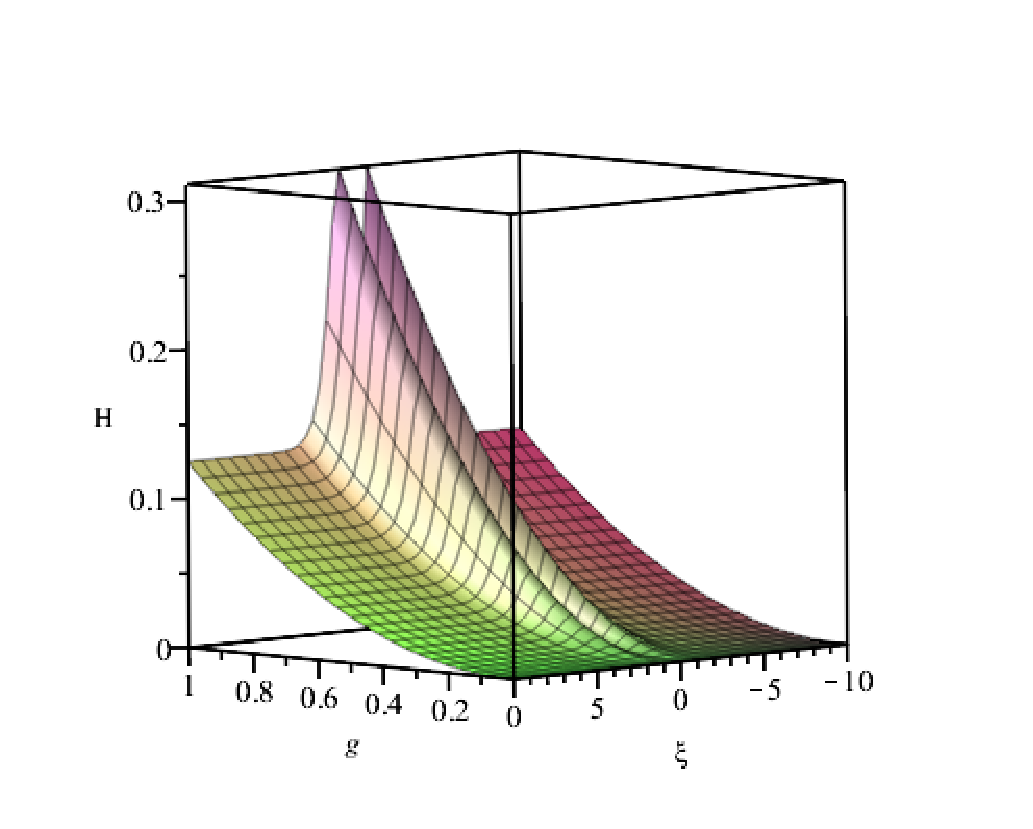}
\caption{\small{The energy density is plotted for $\nu=1$, $b_{1}=1.224$ and $w=0.5$. The figure shows two peaks that display the massive and the massless soliton fields.}}
\end{center}
\end{figure}

 \begin{figure}[ht]
\begin{center}
 \includegraphics[width=6cm]{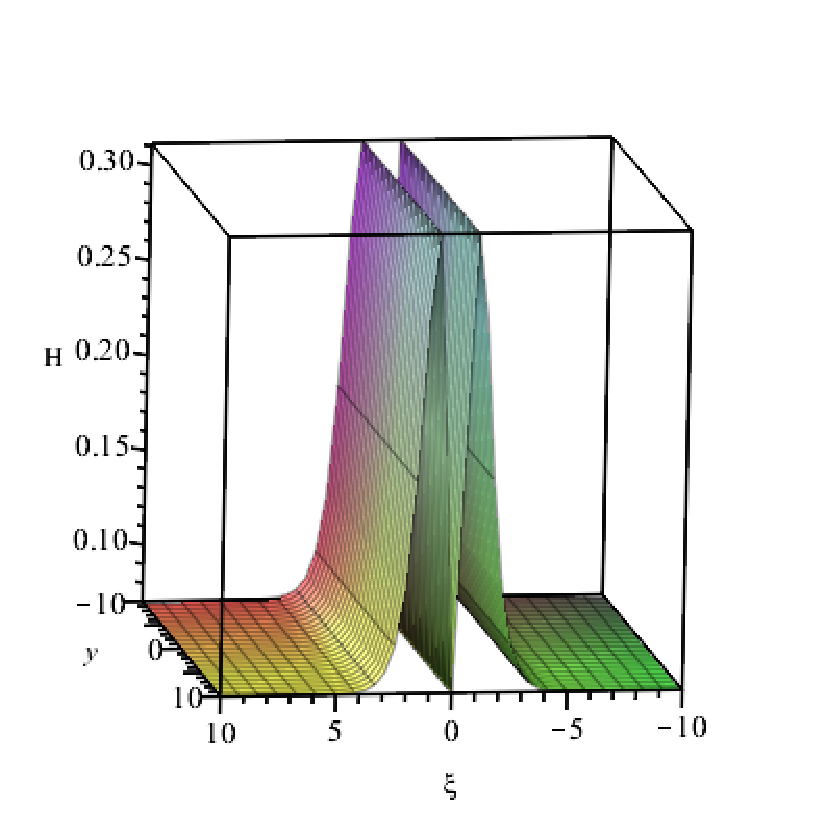}
\caption{\small{
 The energy density is plotted for $\nu=1$, $b_{1}=1.224$, $w=0.5$ and $g=1$.}}
\end{center}
\end{figure}
\subsection{ The Charge - Current density of the model}
Using Eq.(\ref{Ila}), we can find charge density of the model as
\begin{equation}\label{Lu}
 \rho(\xi, w, g)=q\left[-\nu \frac{\partial \psi_{2}(\xi)}{\partial\xi}-\frac{\partial \psi_{2}(\xi) }{\partial\xi}\psi_{1}(\xi)+\frac{\partial \psi_{1}(\xi) }{\partial\xi}\psi_{2}(\xi)\right]\times\left(\frac{\partial \xi}{\partial t}\right)\,,
\end{equation}
by substituting the fields (\ref{Lo}) in the above relation, we have
\begin{eqnarray}
\rho(\xi)&=&-q \nu \left(\sqrt{2\nu^{2}-b_{1}^{2}}sech(\xi)tanh(\xi)\right)-q\left(\sqrt{2\nu^{2}-b_{1}^{2}}sech(\xi)tanh(\xi)\right) \left(-\nu+b_{1}sech(\xi)\right)
\nonumber\\&+& q \left(b_{1}sech(\xi)tanh(\xi)\right)\left(\sqrt{2\nu^{2}-b_{1}^{2}}sech(\xi)\right)=0.\,\label{Lv}
\end{eqnarray}
The above relationship shows that the charge density becomes zero, so the total charge and the electric dipole moment of the system are zero. These results are completely consistent with the neutrality of the produced bosons.
%%%%%%%%%%%%%%%%%%%%%%%%%%%%%%%%%%%%%%%%%%%%%%%%%%%%%%%%%%%%%%%%%%%%%%%%%%%%%%%%%%%%%%%%%%%%%%%%%%%%%%%%%%%%%%%%%%%%%%%%%%%%%%%%%%%%%%%%%%%%%%%%%%%%%%%%%%%%
\section{Conclusions}
In this paper, we have studied the Goldstone boson model which is the simplest model to illustrate SSB under continuous transformation $U(1)$, leading to the appearance of massless and massive spin-zero bosons. The massless bosons have never been observed, but could have many laboratory, astrophysical and cosmological implications. After SSB in the Goldstone boson model, two coupled nonlinear equations appear which have been treated by perturbation theory before, but faced with some problems which were discussed in the text. In this work, we obtained exact solutions for the coupled nonlinear equations by using $\tanh/\mbox{sech}$ method which is an effective technique to solve NPDEs. To apply the method we used a new transformation, namely we changed the coordinates by using a transformation which led us to solitonic solutions for which we chose the localized states. Referring to figure 1, we observed that the solitonic waves are modulated, with the massless solution riding on the massive one. Correspondingly, in the figures representing the energy density, two peaks could be seen that express the existence of two boson fields. It can also be seen that the charge-current density of the model is zero, which again confirms the neutrality of these boson fields.
Since massive bosons can be observed in the laboratory, these solitonic waves may be useful in detecting Goldstone bosons and better understanding dark matter.
Lastly, we note that the approach used in this paper can be extended to examine the model in a curved spacetime scenario.
%These results may be useful in observing massless Goldstone bosons and understanding the  $CP$ violation and  matter-antimatter asymmetry of the Universe. Lastly, we note that the approach used in this paper can be extended to examine the model in a curved spacetime scenario.
\section{Acknowledgment}
The authors would like to thank the anonymous referee for her/his comments and suggestions to improve this paper. Also, the authors are grateful for the valuable comments and suggestions of Prof. Sepangi and Prof. Gousheh to improve the quality of the paper.

%%%%%%%%%%%%%%%%%%%%%%%%%%%%%%%%%%%%%%%%%%%%%%%%%%%%%%%%%%%%%%%%%%%%%%%%%%%%%%%%%%%%%%%%%%%%%%%%%%%%%%%%%%%%%%%%%%%%%%%%%%%%%%%%%%%%%%%%%%%%%%%%%%%%%%%%%%%%

\end{document}